\documentclass[useAMS,usedcolumn,eps,psfig,graphics]{mn2e}
\usepackage{epsfig, graphicx,natbib2}

\def\grtsim{\mathrel{\hbox{\rlap{\hbox{\lower2pt\hbox{$\sim$}}}\raise2pt\hbox{$>$}}}}
\def\lesssim{\mathrel{\hbox{\rlap{\hbox{\lower2pt\hbox{$\sim$}}}\raise2pt\hbox{$<$}}}}

\def\degree{\nobreak\ifmmode{^\circ}\else{$^\circ$}\fi}

\newcommand{\aap}{A\&A}

\newcommand{\apj}{ApJ}

\newcommand{\mnras}{MNRAS}
\newcommand{\nat}{Nat}

\begin{document}

\topmargin -0.5in 

\title[Systematic errors in weighted 2--point correlation functions]{Systematic errors in weighted 2--point correlation functions: An
  application to interaction--induced star formation}
\author[A.~R.~Robaina \& E.~F.~Bell]{Aday R.~Robaina$^{1}$\thanks{E-mail:
    arobaina@mpia.de (ARR)} \& 
Eric F.~Bell$^{2}$\\
\footnotesize\\
$^{1}$Max Planck Institute for Astronomy, Koenigstuhl 17, Heidelberg D--69117,
Germany\\
$^{2}$Department of Astronomy, University of Michigan, 550 Church St., Ann Arbor, MI 48109, USA\\
}

\pagerange{\pageref{firstpage}--\pageref{lastpage}} \pubyear{}

\maketitle

\label{firstpage}

\begin{abstract}
Weighted correlation functions are an increasingly important
tool for understanding how galaxy properties depend on their 
separation from each other.  
We use a mock galaxy sample
drawn from the Millenium simulation, assigning weights
using a simple prescription to illustrate and explore how 
well a weighted correlation function recovers the true
radial dependence of the input weights.
We find that the use of 
a weighted correlation function results in a dilution of the magnitude
of any radial dependence of properties and a smearing out of that
radial dependence in radius, compared to the input behavior.
We present a quantitative discussion of the dilution in the magnitude 
of radial dependence in properties in the special case of a constant
enhancement at $r<r_c$.  
In this particular case where there was 
a SFR enhancement at small radius $r<r_c=35$\,kpc, 
the matching of one member of 
an enhanced pair with an unenhanced galaxy in the same group 
gives an artificial enhancement out to large radius $\sim 200$\,kpc.
We compare this with observations of SFR enhancement from 
the SDSS (Li et al.\ 2008; MNRAS, 385, 1903) 
finding very similar behavior --- a significant
enhancement at radii $<40$\,kpc and a weak enhancement out to more
than 150\,kpc.  While we explore a particular case in this Letter, 
it is easy to see that the phenomenon is general, and precision analyses
of weighted correlation functions will need to account carefully for 
this effect using simulated mock catalogs.
\end{abstract}

\begin{keywords}
galaxies: general --- galaxies:statistics
\end{keywords}

\section{Introduction}

Correlation functions (2-point and higher order ones) have proved to be
powerful statistical tools in order to address the study of the galaxy
clustering \citep[e.g.][]{peebles76,groth,peebles80, dp} and are still widely
used in both local \citep{connolly, eisenstein, masjedi} and high-redshift Universe
\citep{giavalisco, blain}.  Studies of the two point correlation 
function have matured to the point that one can study how galaxies 
populate dark matter halos in detail \citep[e.g.,][]{zehavi}, the 
typical halo masses of galaxy populations as a function of 
redshift (e.g., Lyman breaks - \citealp{giavalisco}), the relative clustering
of different populations (e.g., the tendency of AGN to cluster like 
the massive galaxy population as a whole; \citealp{li06b}), and the use 
of clustering measures on the smallest scale to constrain the merger
history of galaxies (e.g., \citealp{patton}, \citealp{bell}, \citealp{robaina10}).

Furthermore, the correlation function method allows us not only to study the
clustering of the galaxies themselves, but also how some of their properties
are clustered. Weighted correlation functions \citep{boerner} or in a general
sense, marked statistics \citep{beisbart,gott,falten,skibba06,robaina} have been
widely used in the last ten years in order to study how observables
depend on the separation between galaxies.  In particular, 
weighted correlation functions are frequently used to study the 
dependence of star formation rate (SFR) on separation between 
galaxies, in great part to explore the influence of galaxy interactions
on enhancing a galaxy pair's SFR \citep[e.g.][]{li,robaina}.

The goal of this Letter is to explore the application of weighted
correlation functions to study the variation of observables (e.g., SFR, color, 
AGN accretion rate, morphology) as a function of radius.  We briefly 
introduce weighted 2 point correlation functions in \S 2.  We 
then construct a toy model with which we study the behavior of 
the inferred weighted quantities relative to the input behavior (\S 3).  
This toy model is primarily to illustrate some general features
of how weighted correlation functions recover input behavior, and 
we stress that the framework discussed in this Letter applies generally
to any application of weighted correlation function analysis, while 
noting that we choose to present a case that is most directly
analagous to the study of SFR enhancement in close pairs of galaxies.
We show the results of this analysis in \S 4. In \S 5, we briefly
compare with observational results of SF enhancement derived
using the Sloan Digital Sky Survey \citep{li}. In \S 6, we 
present our conclusions. When necessary, we have assumed $H_0=70\,km\,s^{-1}$,
$\Omega_{m0}=0.3$, $\Omega_{\Lambda 0}=0.7$.

\section{Background}

In this work, we explore the possible artifacts that the use of a marked
correlation function could introduce when studying the clustering of galaxy
properties. A full explanation of the methodology followed in this work has
been already presented in \citet{robaina}, and is similar to 
the methodology adopted by \citet{skibba06} and \citet{li}; 
we summarize here the basics of the
method but we refer the reader to those papers for a deeper explanation.

The 2--point correlation function $\xi(r)$ is the {\it excess} probability
of finding a galaxy at a given distance $r$ from
another galaxy:
\begin{equation}{\label{eq:prob}}
dP = n[1+ \xi(r)]dV,
\end{equation}
where $dP$ is the probability of finding a galaxy in volume element 
$dV$ at a distance $r$ from a galaxy, and $n$ is the galaxy number density. 
A simple estimator of the
unweighted correlation function is $\xi(r)\simeq DD/RR-1$, where $DD$ is the histogram
of separations between galaxies and $RR$ is the histogram of separations between
galaxies in a randomly-distributed catalog. In a similar way, 
one can estimate the
weighted correlation function as $W(r)\simeq PP/PP_R-1$, where $PP$ is the
weighted histogram of real galaxies and $PP_R$ the weighted histogram 
of separations from the
catalog with randomized coordinates.

We choose to use an additive weighting scheme (the weight of the pair is the
sum of the weights of individual galaxies) for concreteness 
\citep[e.g.,][]{robaina}, while
noting that a multiplicative weighting would yield
a qualitatively similar result.
Then, we can define the `mark' $E(r)$ as the excess clustering of
the weighted correlation function compared to the 
unweighted correlation function:
\begin{equation}
E(r)=\frac{1+W(r)}{1+\xi(r)}.
\end{equation}

\section{An idealised experiment}

We use \citet{delucia} catalog at $z=0$ derived from 
the Millenium simulation \citep{springel} in order to
study how the enhancement in a physical quantity caused by a galaxy--galaxy
interaction (e.g., a SF enhancement) would be recovered by weighted  2--point correlation function techniques. We
manually assign a weight (we refer to it as the mark) to every
galaxy in the sample, giving a mark=1 to galaxies which are {\it not} closer
than $r_c=35$\,kpc to any other galaxy and mark$=\epsilon$ (with $\epsilon >
1$) to those galaxies which are
in close, 3D pairs with separation $r<r_c$~kpc. 
For concreteness, we consider simulated galaxies 
with stellar masses $M_* > 2.5\times10^{10} M_{\sun}$, 
noting that the conclusions reached in this Letter are generally
applicable, in a qualitative sense.

We now examine how the marks of galaxy pairs relate to the actual behavior
of the enhancement as a function of separation from their nearest neighbor.
The mark is estimated by dividing the weighted correlation function 
by its unweighted counterpart, and recall that the correlation 
function relates every galaxy to every other galaxy in the sample
\footnote{Even in the case in which some criteria for
  pair-matching are imposed,
  like line-of-sight constraints, mass ratio, etc., one particular galaxy will
  be matched with many secondaries at very different separations.}.
The weight is additive, and since every galaxy with a companion closer
than $r_c$ has weight $\epsilon$, the mark of a close pair
is $2\epsilon$.  Yet, the galaxies in this close pair will be matched
also to every other galaxy in the sample.  Therefore, when a galaxy
in the same group or cluster at a distance $r>r_c$ from the enhanced
pair is matched with a galaxy in the pair, the mark of that pair will
be $\epsilon + 1$ (1 being the default weight of non-enhanced galaxies).
We see that a pair with $r>r_c$ will show an enhancement when, in reality,
there is no physical interaction-induced enhancement at that radius.  
As that third galaxy will be
matched with {\it both} galaxies in the neighbor close pair, two pairs with
mark=$\epsilon+1$ will be contributed. Furthermore, imagine now that
there is another real close pair of galaxies placed at several Mpc from the
first close pair, in which both galaxies will also have mark=$\epsilon$. From
matching all those 4 galaxies, the final product will be 6 galaxy pairs
displaying mark=$2 \epsilon$. This will clearly affect both the normalization of
the mark and the recovered value for the enhancement, producing a tail of
false enhancement in the regions where more companions would be found
(representing dense regions of the Universe) and decreasing the enhancement
found at $r<r_c$.

\section{Results}

\begin{figure}
\begin{center}
\psfig{file=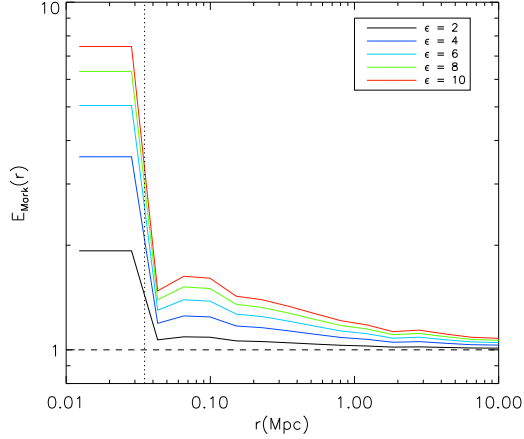,width=7cm,angle=0} 
\caption{Apparent enhancement as a function of distance and "real" enhancement
  $\epsilon$ (which acts only at $r<r_c$\,kpc) computed
for a sample with a minimum stellar mass of $2.5 \times 10^{10} M_\odot$ and different values of $\epsilon$. The vertical dotted line shows
the separation $r_c$ (35 kpc in this case). A tail of artifial enhancement extending to large separations is
produced as an artifact of the weighted correlation function method. The enhancement recovered in the close pairs $r<r_c$\,kpc is reduced respect to $\epsilon$ and the level of reduction is a
strong function of $\epsilon$ (see text and Fig.~\ref{fig:recovered} for more details)}
\label{fig:epsi}
\end{center}
\end{figure}

We show this effect in Fig.~\ref{fig:epsi}. 
Clearly, a relatively weak tail of enhancement is recovered out to
large separations. The amplitude of this tail has a radial dependence, as
close pairs of galaxies tend to be found in dense regions of the Universe
\citep{barton07}. As the magnitude of this
tail depends on the distribution of neighbors as a function of the separation
it will be more relevant for galaxy samples in which the clustering is
stronger (e.g., massive galaxies, or non star-forming galaxies).

\begin{figure}
\begin{center}
\psfig{file=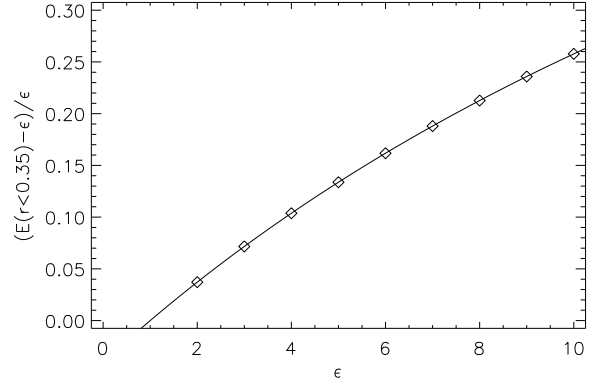,width=8cm,angle=0} 
\caption{Relative error in $E(r<r_C)$, the enhancement recovered by the marked
  correlation function in close pairs as a function of
  the "real" enhancement $\epsilon$ in those pairs. For this example we have used, as in
  Fig. \ref{fig:epsi}
 a lower mass cut of $2.5 \times 10^{10}M_\odot$. {\it Diamonds:} Recovered values
from the method. {\it Solid line:} Expected value using the proper
normalization shown in Eq. \ref{eq:rec}. The error when the intrinsic
enhancement is small is modest; when $\epsilon < 4$ then the discrepancy 
between $E(r<r_c)$ and $\epsilon$ is $<10$\%. }
\label{fig:recovered}
\end{center}
\end{figure}

Also visible in Fig.\ \ref{fig:epsi}
 is the dilution of the recovered
enhancement compared with the actual enhancement $\epsilon$ for 
pairs with $r<r_c$; $E(r<r_c)$. The value of $E(r<r_c)$ is lower than
the "real" enhancement $\epsilon$ by a factor which increases with
$\epsilon$. This effect is better seen in Fig.~\ref{fig:recovered}, where we show
the relative discrepancy between $E(r<r_c)$ and $\epsilon$,
as a function of $\epsilon$. 
In this idealised case, this discrepancy can be exactly recovered
by accounting carefully for the different pairs formed by galaxies 
in the sample.  The relationship between $E(r<r_c)$ and $\epsilon$ is:
\begin{equation}\label{eq:rec_gen}
E(r<r_c)=\frac{\epsilon~ N_{p,tot}}{W_{cp,max} N_{cp,max}+W_{mp}N_{mp}+W_{fp}N_{fp}},
\end{equation}
where $N_{p,tot}$ is the total number
of pairs which can be formed from the galaxy sample\footnote{When performing
  an autocorrelation, the total number of unique pairs would be $N(N-1)/2$, $N$ being the
  number of galaxies in the sample.}, $N_{cp,max}$ is the total number of pairs
which can be formed with galaxies belonging to close pairs\footnote{This is
  {\it not} the same as the number of close pairs, as we already
  explained.},$W_{cp,max}$ is the weight associated with those pairs,
$N_{pm}$ is the number of pairs in which only one galaxy belongs to a
close pair, $W_{mp}$ is the weight associated with them, and $W_{fp}$ and
$N_{p,far}$ are respectively the weight and the number of pairs in which none
of the galaxies belongs to a close pair.

In our particular case of an additive weight, this expression 
reduces to:
\begin{equation}{\label{eq:rec}}
E(r<r_c)=\frac{2\epsilon}{(f^2+f)(\epsilon-1)+2},
\end{equation}
where $f$ is the fraction of galaxies in close pairs. The degree of clustering
of the sample is reflected in the value of $f$, 
so this expression is valid under different
clustering conditions. For the purposes of this work, we calculate $f$ {\it
  directly} from the mock catalogue, but real galaxy surveys lack of accurate
3D information. It is common to calculate $f$ from the
inferred real space correlation function by integrating Eq.\ 1 out to $r_c$
\citep{masjedi, bell}.  In their analysis, in the limit
of small $r_c$, and if the correlation function is parametrised as a power law
$\xi(r)=(r/r_0)^{-\gamma}$, then:
\begin{equation}
P(r<r_c)=f=\int_{0}^{r_c}n[1+\xi(r)]dV
\end{equation}

\begin{equation}{\label{eq:prob2}}
f \simeq \frac{4\pi n}{3-\gamma}r_0^\gamma r_c^{3-\gamma}.
\end{equation}

It is worth noting that in the above 
example we have studied the simple case in which
the enhancement is present only in close galaxy pairs, with the enhancement
represented by a step function. When applying weighted correlation functions to
more complex problems, like those involving clustering of the mass or colour,
the function describing the behavior of the weight on separation  
would be much more complex. In that case,
an expression for the behavior of the weight as a function of separation
will have to be derived on a case-by-case basis and matched with the data.
Yet, even in that more complex case, the underlying problem is 
very similar: the magnitude of any radial dependence in properties
will be diluted and smeared out in radius by the use of weighted
2 point correlation function methods.
 
\section{An example application to observations}

In order to test the relevance of this analysis to the real Universe, 
we compare our predictions with a 
well-established phenomenon: the
enhancement of the star formation rate (SFR) in galaxy interactions. This
observable has two obvious advantages. Firstly, there are a number of works in
which this enhancement has been studied \citep{barton,lambas,li,robaina}.
Second, the SFR is expected to be enhanced only at scales at which
galaxy-galaxy interactions are relevant; beyond that scale star formation
is not only not expected
to be enhanced, but should be depressed because of 
the well known SFR-density anticorrelation \citep[e.g., ][]{balogh}. 
From the above mentioned works we choose to compare with
\citet{li} for three reasons: a) they use marked statistics, b) their
large sample allowed an accurate estimate of enhancement to be made, 
and c) SDSS clustering has been shown to be similar to the
one present in the \citet{delucia} mock catalogue from the Millenium Simulation
in the local Universe \citep{springel}.

\begin{figure}
\begin{center}
\psfig{file=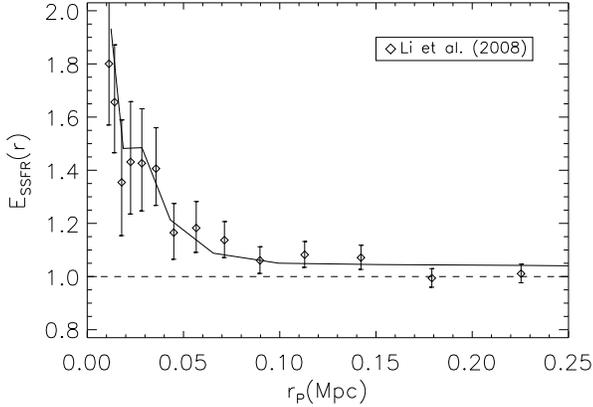,width=8cm,angle=0} 
\caption{Specific SFR enhancement in (massive) galaxy pairs as function
  of the projected 
  separation as measured by \citet{li} (diamonds) and our prediction
  including the tail of artificial enhancement out to several hundred
  kiloparsecs (solid line). In this example a value of $\epsilon=2$ has been
used for galaxies in pairs with $r_P< 15$ kpc and $\epsilon=1.5$ for galaxies
in pairs with $15 < r_P< 40$ kpc. The galaxy sample has been selected to be consistent with the massive
sample in \citet{li}.}
\label{fig:sfr}
\end{center}
\end{figure}

Real galaxy surveys, even spectroscopic surveys, have no access to the 
real space separation of galaxies.  \citet{li} used a projected
correlation function $w(r_P)$ to circumvent this difficulty, where
the projected correlation function is related to the 3D correlation function
via:
\begin{equation}
w(r_p) = \int^{\infty}_{- \infty} \xi ( [r_p^2 + \pi^2]^{1/2}) d \pi,
\end{equation}
where $\pi$ is the coordinate along the line of sight, and $r_p$ is the 
projected separation transverse to the line of sight.  
We use for this exercise galaxies more massive than $3\times
  10^{10}M_\odot$ in order to match the selection citeria in
  \citet{li}. Moreover, they did not use an additive weight but used the SSFR of
  the primary galaxy as the weight of the pair. We also use such a scheme here
  to perform our weighted analysis in the simulation.  \citet{li} calculated the cross-correlation between a subsample of galaxies which are forming stars (primaries)
and all the galaxies in the sample (secondaries). As we lack of such
information we run a correlation using all the galaxies as both primaries and
secondaries. As previously, we assign an average enhancement to all
  galaxies found physically in very close pairs, but in order to mimic the the
  pair selection in \citet{li}, who cross-correlate a sample of
  spectroscopically defined star forming galaxies with a photometric catalog
  of all galaxies above the stellar mass limit, we run the correlation function
  selecting galaxy pairs with ``line-of-sight'' separations of less than 100
  Mpc. Our results are not sensitive to this choice of maximum separation;
  correlations between galaxies on scales larger than 100 Mpc are extremely
  weak, in comparison to the strong clustering on $<1 Mpc$ scales. We choose to model the data with 
a constant enhancement $\epsilon=1.8$ at $r<r_c$, with $r_c = 35$\,kpc,
  Motivated by the star formation enhancement observed in galaxy samples
  selected in a similar manner at different redshifts
\citep{li, robaina} we choose to model the data with $\epsilon =2$ for
galaxies in pairs with separations $r_P<15$ kpc and $\epsilon=1.5$ for those
in pairs with $15<r_P<40$ kpc. We also neglect any environmental suppression of star formation at separations 
$r>r_c$ \citep{barton, balogh}.
These are clearly oversimplifications, as the real dependence of
enhancement (and suppression at large radii) 
on separation will be considerably more complex.  
Yet, this simple model suffices to illustrate the recovered enhancement 
signature expected from a model in which SF is enhanced only at small
radii.

Notwithstanding these limitations, we compare the results of our 
simple model with the data in Fig.\ \ref{fig:sfr}.  Strikingly, we find that 
the tail of enhanced SF out to $\sim$200kpc seen in the data may, in great 
part, be a reflection of the use of marked correlation functions statistics
to explore the radial dependence of SF enhancement in galaxies.
This has direct relevance in the interpretation of
the results from \citet{li}.  
If one argued that the enhancement at $\sim 100$\,kpc (or much of it) was 
real, one would need to fulfil two criteria to produce such an effect.
Firstly, assuming that the triggering event is the first pass, 
one would need an enhancement lifetime of at least 300Myr (longer than 
the internal dynamical time)
for typical orbital velocities of 300km/s or less. Secondly, a significant
fraction of the secondaries would need to have near-radial orbits 
in order to produce such an enhancement.   
If, as we suggest instead, the enhanced SF at $\sim 100$\,kpc
is an artifact of the use of the 2 point correlation function, then 
one would argue that enhancement happens only for close pairs 
and shorter interaction-induced SF timescales and a greater diversity of
orbits would be permitted.  While developing a model that 
realistically reproduces the data is beyond the scope of this Letter, 
one can clearly see that this effect needs to be accounted for 
in order to robustly interpret the behavior of marked correlation 
functions.

\section{Conclusions}

Weighted correlation functions are an increasingly important
tool for understanding how galaxy properties depend on their 
separation from each other.  
We use a mock galaxy sample
drawn from the Millenium simulation, assigning weights
using a simple prescription to illustrate and explore how 
well a weighted correlation function recovers the true
radial dependence of the input weights.
We find that the use of 
a weighted correlation function results in a dilution of the magnitude
of any radial dependence of properties and a smearing out of that
radial dependence in radius, compared to the input behavior.
We present a quantitative discussion of the dilution in the magnitude 
of radial dependence in properties in the special case of a constant
enhancement $\epsilon$ for pairs separated by $r<r_c$.  
In this particular case  
the matching of one member of 
an enhanced pair with an unenhanced galaxy in the same group 
gives an artificial enhancement $\sim 0.1 \epsilon$ out to 
large radii $\ga 5r_c$, 
and matches of one member of an enhanced pair with a member of another 
very distant enhanced pair pulls down the value of the recovered enhancement, 
with the discrepancy between the input and recovered enhancement being 
a function of the fraction of galaxies in close pairs and the 
value of the input enhancement. This systematic error is $<10\%$ for
enhancements $\epsilon < 4$, but precision measurements should account for
this effect.
We compare these results with observations of SFR enhancement from 
the SDSS \citet{li}, finding very similar behavior --- a significant
enhancement at radii $<40$\,kpc and a weak enhancement out to more
than 150\,kpc, lending credibility to the notion that weak 
enhancement in SFR seen out to large radii is an artifact of the
use of weighted correlation function statistics.  
While we explored a particular case in this Letter, 
it is easy to see that the phenomenon is general.

Given this difference between input weights and those recovered by 
the weighted 2 point correlation function, one might ask if 
one shouldn't use a different method to 
explore radial trends in observables.  We would argue that most different
methods boil down to weighted 2 point correlation functions implicitly anyway, 
and that one is stuck at least at the qualitative level 
with the differences between input and recovered
weights that we have discussed above.  For example, partnering projected
pairs into different 'pairs' (i.e., not matching every galaxy with every 
other galaxy) suffers from two drawbacks: this is still a projected 
analysis, and many projected close pairs will be separated by significant
distances along the line of sight; and second, one may choose the wrong 
galaxy to partner with, a particularly acute issue for triplets or
groups of galaxies.  One can see that such a method will suffer from 
a similar supression of enhancement from the inclusion of non-pairs in 
the pair sample; of course, radial smearing is not possible in such a case, 
as there is only one radial bin.  We conclude that those wishing to 
quantitatively
analyze weighted correlation functions (or related observables) 
will need to account carefully for 
this effect using an analysis of simulated mock catalogs.

\section*{Acknowledgments}

We thank Cheng Li for sharing with us the electronic version of the results of
their work.  A.\ R.\ R.\ acknowledges the Heidelberg--International Max Planck Research School program.

\end{document}